\documentclass[final]{ws-ijmpe}

\usepackage{amssymb}

\def\beh{\begin{equation}}
\def\eeh{\end{equation}}

\begin{document}

\markboth{Pawel Danielewicz and Jenny Lee}{Symmetry Energy in Nuclear Surface}

%
\catchline{}{}{}{}{}
%

\title{SYMMETRY ENERGY IN NUCLEAR SURFACE
}

\author{\footnotesize PAWEL DANIELEWICZ\footnote{danielewicz@nscl.msu.edu}\,\, AND JENNY LEE\footnote{lee@nscl.msu.edu}}

\address{National Superconducting Cyclotron Laboratory and\\
Department of Physics and Astronomy, Michigan State University,
\\
East Lansing, Michigan 48824, USA}

\maketitle

\begin{history}
\received{(received date)}
\revised{(revised date)}
\end{history}

\begin{abstract}
Interplay between the dependence of symmetry energy on density and the variation of nucleonic densities across nuclear surface is discussed.  That interplay gives rise to the mass dependence of the symmetry coefficient in an energy formula.  Charge symmetry of the nuclear interactions allows to introduce isoscalar and isovector densities that are approximately independent of the magnitude of neutron-proton asymmetry.
\end{abstract}

\section{Introduction}

The utility of nuclear symmetry energy in assessing nuclear properties is a consequence of the charge symmetry of nuclear interactions, an invariance of those interactions under neutron-proton interchange.  A broader symmetry is the charge invariance, an invariance under rotations in neutron-proton space.  In the following, we shall explore different consequences of both types of invariance on nuclear characteristics, as well as interdependencies between those consequences.  Much of the discussion is self-contained; additional technicalities can be found in Refs.\cite{Danielewicz:2008cm,Danielewicz:2003dd}.

One consequence of the charge symmetry is that isoscalar quantities can be identified within a nuclear system, that do not change under the proton-neutron interchange.  These include the nuclear part of net energy and the net nucleon density, $\rho (r) = \rho_n (r) + \rho_p (r)$.  When expanded in relative asymmetry, $\eta=(N-Z)/A$, an isoscalar quantity, $F(\eta)$, contains even powers of $\eta$ only,
$
F(\eta) = F_0 + F_2 \, \eta^2 + F_4 \, \eta^4 + \ldots
$
Due to the lack of a linear term, an isoscalar quantity depends weakly on asymmetry.  Notably, for the expansion to apply, a nuclear quantity such as energy must be smoothed out to suppress shell and pairing effects.  Apart from the isoscalar, isovector quantities may be identified, that change sign under the $n \leftrightarrow p$ interchange.  An example is the neutron-proton density difference, $\rho_{np}(r) = \rho_n(r) - \rho_p(r)$.  When expanded in~$\eta$, an isovector quantity, $G(\eta)$, contains odd powers of $\eta$ only,
$
G(\eta) = G_1 \, \eta + G_3 \, \eta^3 + \ldots
$
Dividing an isovector quantity by~$\eta$, or by another isovector quantity, yields an isoscalar quantity, weakly dependent on~$\eta$: $G/\eta = G_1 + G_3 \, \eta^2 + \ldots$.  When considering the charge invariance, isovector quantities need to transform in a covariant fashion under a broader class of $n$-$p$ transformations.

Within the realm of consequences of charge-symmetry, we will be interested in the interplay of the density-dependence of symmetry energy in uniform matter and the characteristics of nucleonic densities across nuclear surface.  We shall test our qualitative assertions within the Skyrme-Hartree-Fock (SHF) calculations of half-infinite nuclear matter that is the simplest nuclear system with a surface.  In this context, we shall further consider mass-dependence of the symmetry coefficient in an energy formula.

\section{Symmetry Energy}

On account of the charge symmetry, the expansion of energy per nucleon of uniform matter, in asymmetry, takes the form
\beh
\label{eq:EA}
\frac{E}{A}(\rho_n, \rho_p) = \frac{E_0}{A} (\rho) + { S(\rho)} \left( \frac{\rho_n-\rho_p}{\rho} \right)^2 + \ldots
\eeh
The leading term in this expansion is the energy per nucleon of symmetric matter.  The remainder, the leading term of the remainder, and the coefficient $S$ in the remainder are all called the symmetry energy in literature.  Under most circumstances, higher-order terms, beyond the first two on the r.h.s., play little role in the net energy.  Correspondingly, the knowledge of two functions of density, $\frac{E_0}{A}(\rho)$ and $S(\rho)$, allows, in practice, to describe the nuclear energy in a wide density and asymmetry range. E.g.\ in pure neutron matter, the nuclear energy is $\frac{E}{A} \simeq \frac{E_0}{A} + S$.  The symmetry energy is customarily expanded around the normal density~$\rho_0$, in a manner similar to the energy of symmetric matter, $S(\rho) = a_a^V + \frac{L}{3} \frac{\rho- \rho_0}{\rho_0} + \ldots$  Because the energy of symmetric matter minimizes at $\rho_0$, the constant $L$ determines the nuclear contribution to net pressure in the vicinity of~$\rho_0$, that is important for the structure of neutron stars\cite{Lattimer:2006xb}, $P \simeq \rho^2 \, \frac{{\rm d}S}{{\rm d}\rho} \simeq  L \, \frac{ \rho^2}{3\rho_0}$.

In the Bethe-Weizsacker formula, expressing the energy of a nucleus in terms of nucleon numbers, the charge symmetry constraints the symmetry term to a~quadratic form in asymmetry:
\beh
\label{eq:EBW}
{ E = - a_V \, A + a_S \, A^{2/3}
+ a_C \, \frac{Z^2}{A^{1/3}} +} { a_a (A)} \,{\frac{(N
-  Z)^2}{A} } { + E_{mic}} \, .
\eeh
The symmetry coefficient $a_a$ can principally depend on mass number~$A$, rather than having a fixed value ($a_a \equiv a_a^V$) that is usually assumed.  We shall explore a~possibility of the mass dependence below.

Due to its quadratic dependence on asymmetry, the nuclear contribution to the macroscopic energy of a nucleus,
\beh
\label{eq:Emac}
E = -a_v \, A + a_s \, A^{2/3} + \frac{a_a}{A} \, (N-Z)^2 = E_0(A) + \frac{a_a(A)}{A} \, (N-Z)^2 \, ,
\eeh
is analogous\cite{Danielewicz:2003dd} to the energy of a~capacitor, $E = E_0 + Q^2/2C$, with the asymmetry $N-Z$ being analogous to the charge $Q$ and with the ratio $A/2a_a$ being analogous to the capacitance~$C$.  The asymmetry chemical potential,
\beh
\mu_a = \frac{\partial E}{\partial (N-Z)} = (N-Z) \, \frac{2a_a(A)}{A} \,  ,
\eeh
is, in this context, analogous to the electric potential $V$ across capacitor terminals, $V = Q/C$.  Turning back to uniform matter, we have
\beh
\label{eq:mua_uni}
\mu_a = \frac{\partial E}{\partial (N-Z)}
=  \rho_{np} \,  \frac{2 S(\rho)}{\rho} \, ,
\eeh
and the ratio $\rho /2 S$ can be identified as capacitance per unit volume, cf.~(\ref{eq:EA}).  In connection to the electrostatic analogy, we may note that, for connected independent capacitors, the net charge partitions itself in proportion to individual capacitances and the capacitances, otherwise, add up.

\section{Nuclear Densities}

We have mentioned that the net nucleon density $\rho(r) = \rho_n(r) + \rho_p(r)$ is isoscalar and, as such, should weakly depend on $(N-Z)$ for a given~$A$.  We will be ignoring here the Coulomb effects that may be accounted for in terms of corrections.  The density difference, $\rho_{np}(r) = \rho_n(r) - \rho_p(r)$, is isovector but the renormalized density $A \, \rho_{np} (r)/(N-Z)$ is isoscalar in nature.  Deficiency of the net relative asymmetry, as a renormalizing factor, is that it refers to global system characteristics. Quantitatively similar factor, that pertains to local properties only, is $\mu_a/2 a_a^V$.  With this, we introduce asymmetric density, a~counterpart to the net density, as
\beh
\rho_a (r) = \frac{2 a_a^V}{\mu_a} \left[\rho_n(r) - \rho_p(r) \right]  \, .
\eeh
The asymmetric density serves as an isoscalar formfactor for the isovector density.  With the specific normalization, the asymmetric density approaches normal density $\rho_0$ when the net density itself approaches~$\rho_0$, in a weakly nonuniform system.

As a consequence of charge symmetry, the nucleonic densities may be represented in terms of two densities, $\rho$ and $\rho_a$, that each depends weakly on asymmetry\cite{Danielewicz:2008cm},
\beh
\rho_{n,p}(r) = \frac{1}{2} \big[ \rho(r) \pm \frac{\mu_a}{2 a_a^V} \rho_a(r) \big] \, .
\eeh
It is common to approximate the net nuclear density in terms of a Fermi shape
$
\rho(r) = {\rho_0}/[{1 + {\rm exp}( \frac{r - R}{d} )}]
$,
where $R= r_0 \, A^{1/3}$.  Obviously, it is interesting to ask what the corresponding features of $\rho_a$ are.  It turns out that those features are simultaneously related to $a_a(A)$ and $S(\rho)$.

Thus, the net capacitance for asymmetry may be written as
\beh
\label{eq:capA}
\frac{A}{2 \, a_a(A)} = \frac{N-Z}{\mu_a} = \int {\rm d}r \, \frac{\rho_{np}}{\mu_a}
= \frac{2}{a_a^V} \int {\rm d}r \, \rho_a(r)  \, .
\eeh
It follows then that the generalized asymmetry coefficient may be obtained by integrating the asymmetry density over volume.  Upon subtracting and adding the isoscalar from the isovector density, we further find
\beh
\label{eq:aaA_exp}
\frac{A}{a_a(A)} = \frac{1}{a_a^V} \int{\rm d}^3 r \, \rho (r) + \frac{1}{a_a^V} \int {\rm d}^3 r \, (\rho_a - \rho) (r)
 \simeq \frac{A}{a_a^V} + \frac{A^{2/3}}{a_a^S} \, ,
\eeh
In the last step, we have noted that the two densities are close to $\rho_0$ within the nuclear interior; thus, the last integration is concentrated to the surface region for a sufficiently large system, with the result being then proportional to~$A^{2/3}$.  The surface modifies the capacity of a system for asymmetry, as compared to that expected for a uniform system at normal density.  For large systems, the~modification may be described in terms of the surface symmetry coefficient~$a_a^S$.

Following (\ref{eq:mua_uni}), the neutron-proton density difference in a uniform system is
\beh
\label{eq:rhonp_loc}
\rho_{np} = \mu_a \, \frac{\rho}{2 \, S(\rho)} \, ,
\eeh
and the isovector density is then
\beh
\label{eq:rhoa_loc}
\rho_a = \frac{2  a_a^V \, \rho_{np}}{\mu_a} = \frac{a_a^V \, \rho}{S(\rho) } \, .
\eeh
Given the short-range nature of nuclear interactions, these results should hold for weak nonuniformities, with the densities pertaining then to a~specific location.  In the context of (\ref{eq:rhoa_loc}), we can rewrite the result (\ref{eq:capA}) for the system capacitance as
\beh
\label{eq:cap_loc}
\frac{A}{2 \, a_a(A)} \simeq \int {\rm d}^3 r \, \frac{a_a^V \, \rho (r)}{2 S(\rho(r))} + \ldots \, .
\eeh
Here, the first r.h.s.\ term represents the regions where the local approximation (\ref{eq:rhoa_loc}) holds, and the dots represent the remainder.  In (\ref{eq:cap_loc}), the net capacitance of the system emerges as an integral over independent capacitances at different localities; in (\ref{eq:rhonp_loc}) the asymmetry is seen to partition itself in proportion to the local capacitance.  In the context of (\ref{eq:rhoa_loc}) and (\ref{eq:cap_loc}), one can observe that dropping of symmetry energy in the surface region should act to enhance asymmetric density there and to increase the net system capacitance for asymmetry.  The relative increase in net capacitance, cf.~(\ref{eq:aaA_exp}), should be more pronounced in light systems where the surface plays a larger role than in heavy systems.

\section{Half-Infinite Matter in Skyrme-Hartree-Fock Calculations}

We next verify the validity of our qualitative claims by examining the features of half-infinite matter in the Skyrme-Hartree-Fock (SHF) calculations\cite{Danielewicz:2008cm,Farine:1980}.  The direction of uniformity is taken as~$z$, with the vacuum and matter positioned at the positive and negative ends of the $z$-axis, respectively.  The wavefunctions are taken in the form $\Phi({\bf r}) = \phi(z) \, {\rm e}^{i {\bf k}_\perp \cdot {\bf r}_\perp}$, the wavevector space is discretized and we solve the equations for individual wavefunctions,
\beh
-\frac{{\rm d}}{{\rm d} z} \frac{\hbar^2}{2 m^*(z)} \frac{{\rm d}}{{\rm d} z} \phi(z) + \left( \frac{\hbar^2\, k_\perp^2}{2 m^*(z)} +  U(z) \right)  \phi(z) = \epsilon({\bf k}) \, \phi(z) \, ,
\eeh
where $m^*$ and $U$ are obtained through self-consistency.

\begin{figure}[th]
\centerline{\psfig{file=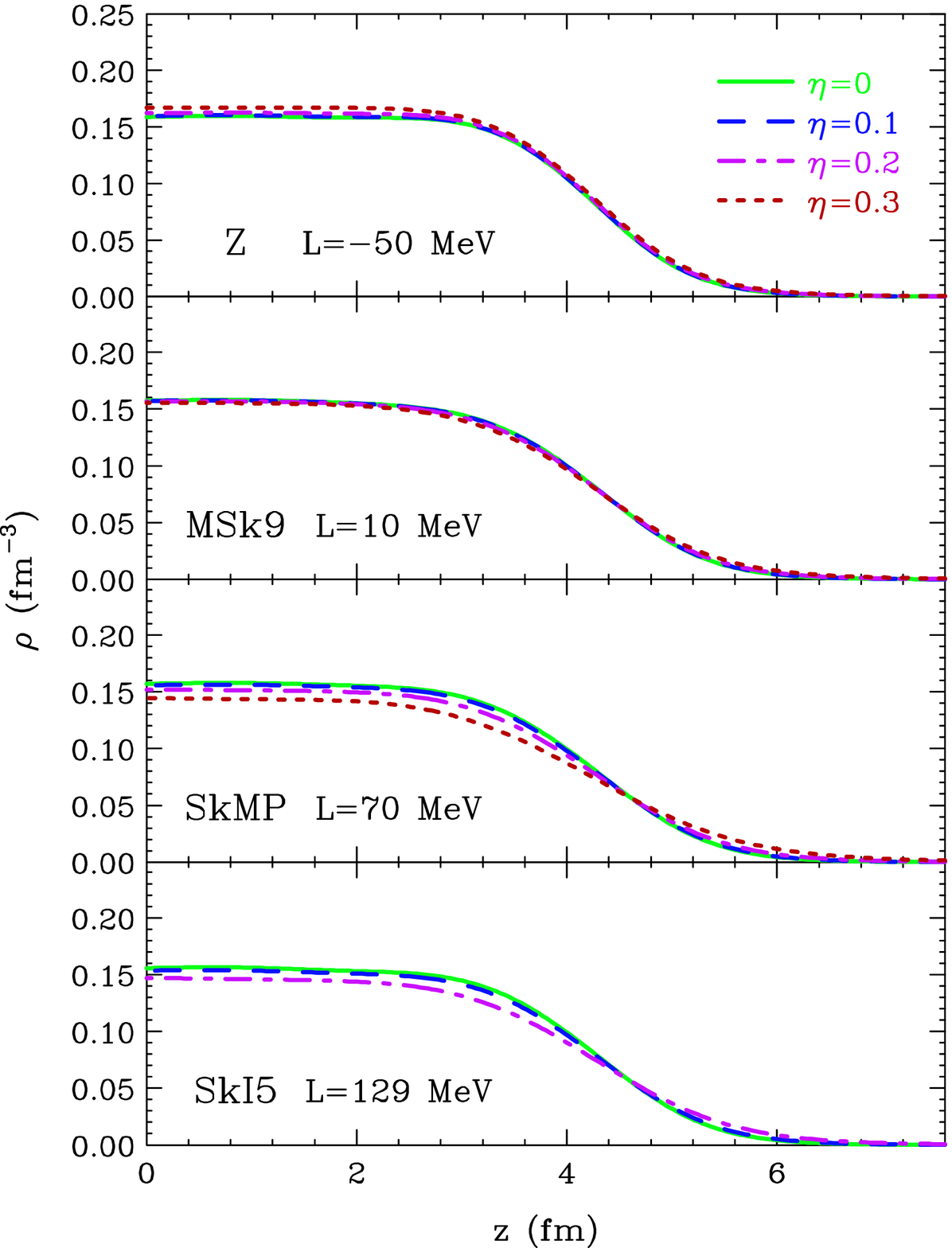,width=6cm} \hspace*{.5em} \psfig{file=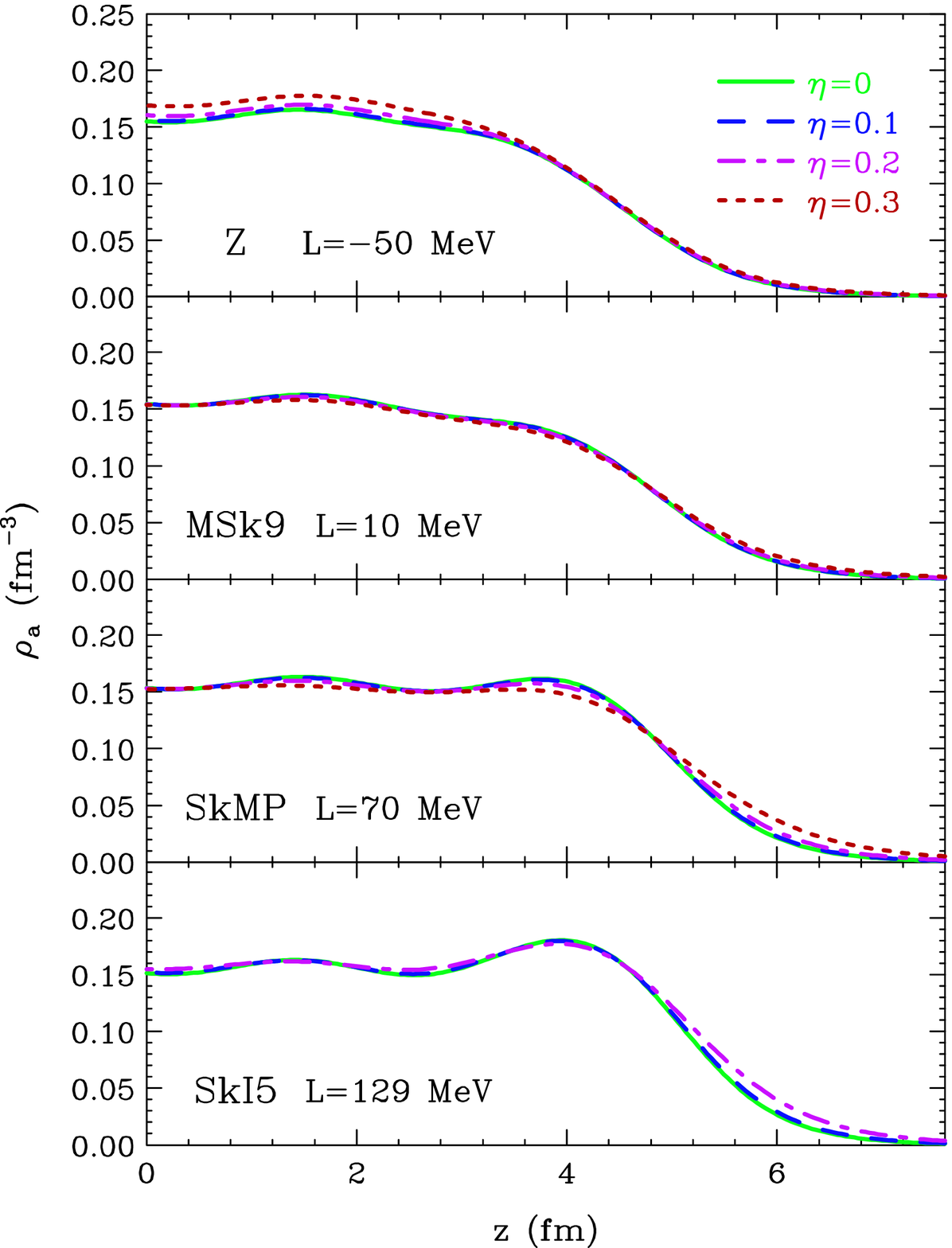,width=6cm}}
\vspace*{8pt}
\caption{Isocalar (left) and isovector (right) density profiles at different asymmetries in half-infinite nuclear matter from SHF calculations, for sample interactions considered in Ref.\protect\cite{Danielewicz:2008cm}.}
\label{fig:density}
\end{figure}

We have claimed that the isoscalar and isovector densities should change little with asymmetry.  The density profiles for the nuclear matter at different asymmetries, for different interactions in SHF calculations, are shown in Figure~\ref{fig:density}.  It is apparent that indeed either density changes little with~$\eta$.  On the other hand, the isovector density changes significantly with interaction.

\begin{figure}[th]
\centerline{\psfig{file=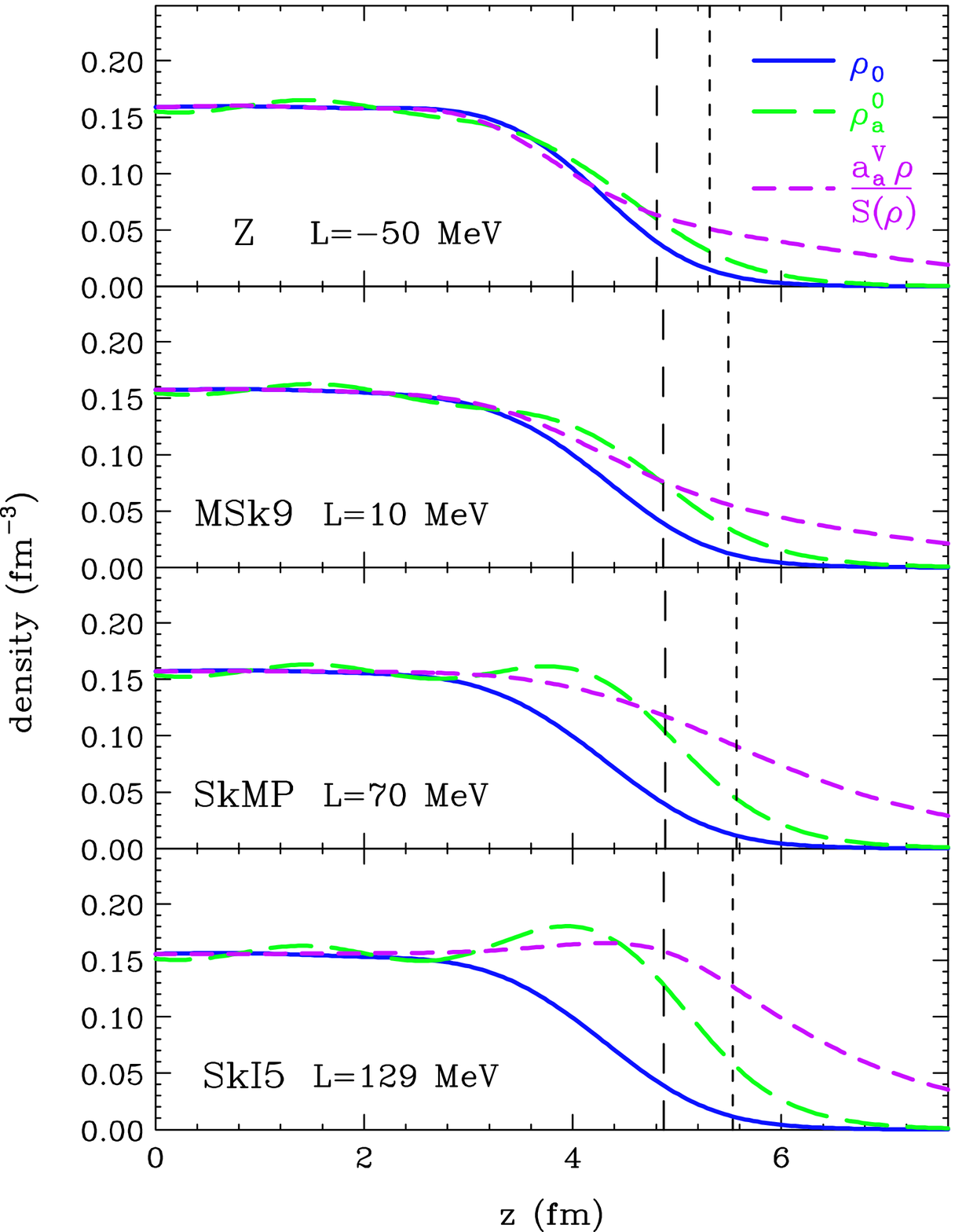,width=6cm} \hspace*{1em} \psfig{file=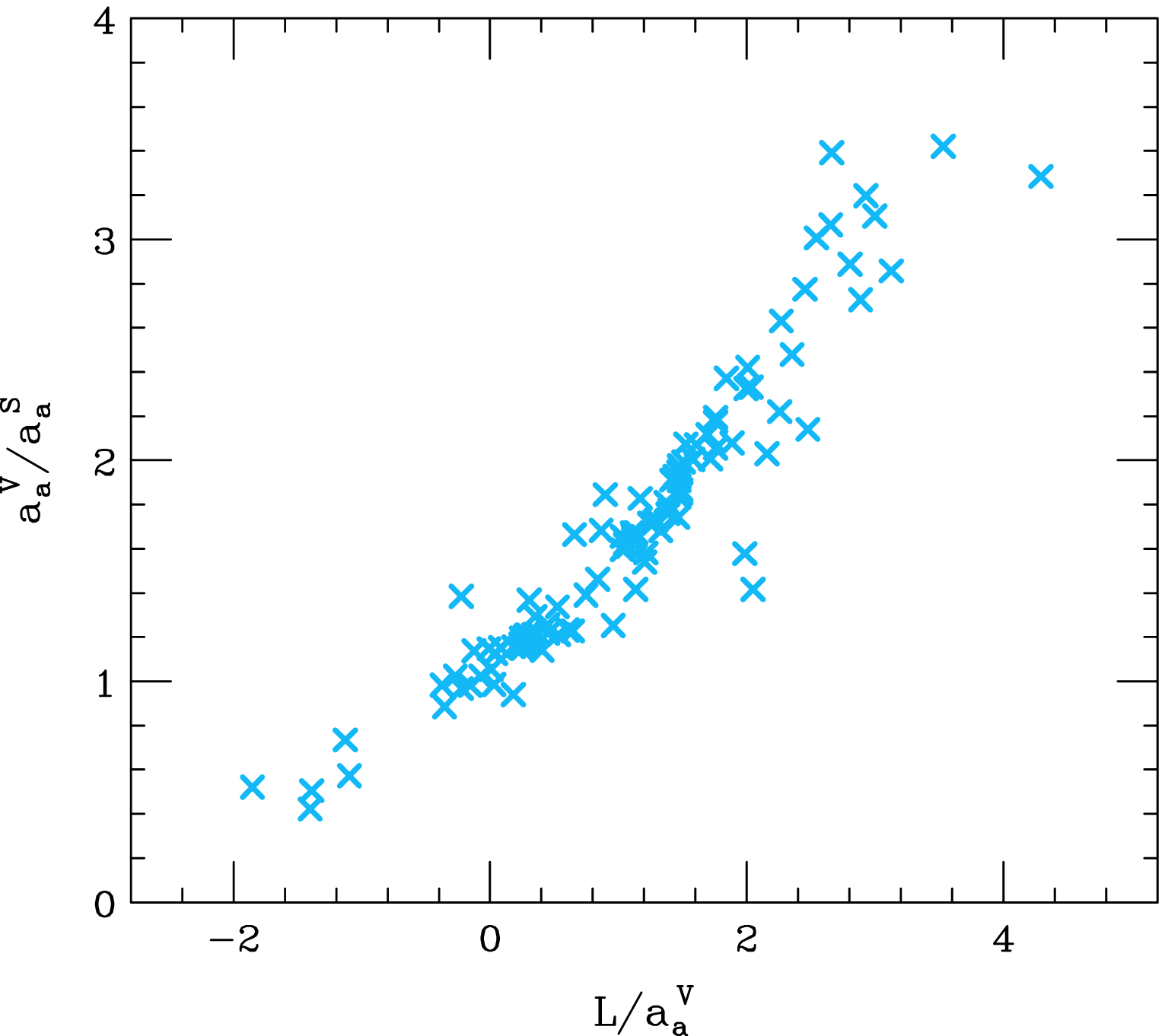,height=6.3cm,width=6cm} }
\vspace*{8pt}
\caption{Left panel shows the comparison of isoscalar and isovector densities, and of local approximation to the isovector density, in symmetric half-infinite nuclear matter from the SHF calculations, for sample interactions.  The~longer- and shorter-dashed vertical lines, for a specific interaction, indicate, respectively, the~location where the net density is equal to the quarter of normal density and the location of a classical return point for the Fermi wavevector directed along the $z$-axis.  Right panel shows the ratio of symmetry energy coefficients $a_a^V/a_a^S$, for different Skyrme interactions\protect\cite{Danielewicz:2008cm}, plotted vs the slope parameter $L$ of the symmetry energy, scaled with the value of symmetry energy $a_a^V$.}
\label{fig:denscomp}
\end{figure}

Figure \ref{fig:denscomp} shows next the comparison of isoscalar and isovector densities for the different interactions.  The~larger the value of slope parameter~$L$ for the symmetry energy, the farther is the isovector density pushed out relative to the isoscalar density. This is consistent with the expectation based on the local approximation of Eq.~(\ref{eq:rhoa_loc}).  The local approximation is also tested in Fig.~\ref{fig:denscomp}.  Up to the Friedel oscillations\cite{Friedel54}, with a wavelength of $\lambda = \pi/ k_F \simeq 2.36 \, {\rm fm} $, the isovector density follows the local approximation down to the net density equal to about a quarter of normal density.  An analysis\cite{Danielewicz:2008cm} following the WKB approximation shows that the local approximation may hold in the classically allowed region; in that respect, the quarter of normal density appears then to represent a typical classical return point.

The farther the isovector density extends outside of the isoscalar density, with the larger~$L$, the more significant is the modification of system capacitance by the surface, and the smaller is the surface symmetry coefficient $a_a^S$, cf.~(\ref{eq:capA}) and~(\ref{eq:aaA_exp}).  The correlation between the ratios $a_a^V/a_a^S$ and $L/a_a^V$ is shown, for the Skyrme interactions, in the left panel of Fig.~\ref{fig:denscomp}.  Typical value of the volume coefficient for the interactions is $a_a^V \sim 30 \, {\rm MeV}$.

\section{Constraints from Isobaric Analog States}

We now turn to the constraints on symmetry energy stemming from an application of the energy formula~(\ref{eq:EBW}).  Unfortunately, in practice, competition between different physics terms within an energy formula makes it difficult to learn about the mass dependence of the symmetry coefficient, by fitting directly the formula to the ground-state nuclear energies\cite{Kir08}.  However, the unwanted competition may be practically eliminated by exploiting the charge invariance of nuclear interactions and generalizing the macroscopic energy formula to the lowest states of a given net isospin within a nucleus\cite{Janecke:2003}.  The~generalization amounts to the replacement of the symmetry term in Eq.~(\ref{eq:EBW}):
\beh
a_a(A) \, \frac{(N-Z)^2}{A} = 4 \, a_a(A) \,  \frac{T_z^2}{A} \Rightarrow 4 \, a_a(A) \, \frac{T(T+1)}{A} \, .
\eeh
In the ground state of a nucleus, the isospin takes on the lowest possible value $T=|T_z| = |N-Z|/2$.
This replacement absorbs the so-called Wigner term from~$E_{\rm mic}$.  Following the formula generalization, it becomes possible to deduce the symmetry coefficient within a single nucleus, by using excitation energies to the isobaric analog states (IAS) representing the ground states of neighboring nuclei\cite{Danielewicz:2004et}, with
\beh
\label{eq:AaaA}
\Delta E = 4 \, a_a(A) \, \frac{\Delta \big( T(T+1) \big) }{A} + \Delta E_{mic} \, .
\eeh

\begin{figure}[th]
\centerline{\psfig{file=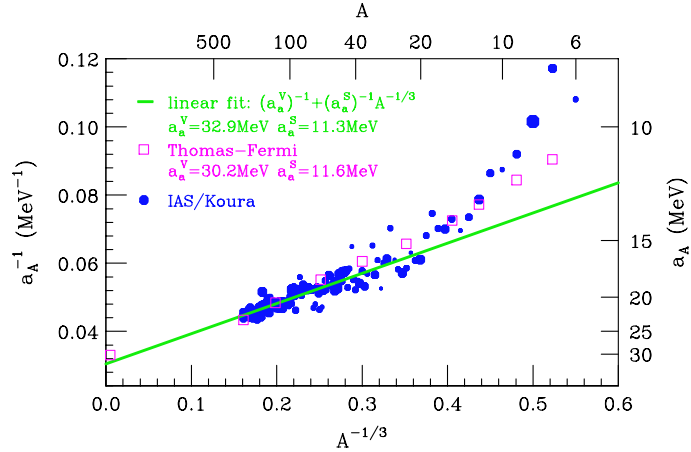,width=9cm} }
\vspace*{8pt}
\caption{Inverse of the symmetry coefficient, on the left scale, as a function of the inverse cube root of mass number, on the bottom scale.  The filled circles represent the coefficients extracted from IAS excitation energies, with the microscopic corrections applied.  The line shows a fit to the IAS results for $A > 20$.  The squares represent the coefficients from a Thomas-Fermi model that best describes the data.}
\label{fig:pa3}
\end{figure}

We use the tabulated energies of isobaric analog states\cite{Antony:1997} and microscopic corrections to energies by Koura {\em et al.}\cite{Koura:2005} to deduce the symmetry coefficients for individual nuclear masses.  The mentioned corrections include deformation effects.  Deduced coefficient values range from $a_a \sim 9 \, {\rm MeV}$, for light $A < 10$ nuclei, to $a_a \sim 22.5 \, {\rm MeV}$, for $A > 200$.  Figure~\ref{fig:pa3} shows inverse coefficient values plotted against inverse cube root of mass number.  For $A > 20$, the value systematic is approximately linear, as expected from Eq.~(\ref{eq:aaA_exp}).  A fit with the r.h.s.\ of Eq.~(\ref{eq:aaA_exp}) produces coefficient values of $a_a^V = 32.9 \, {\rm MeV}$ and $a_a^S = 11.3 \, {\rm MeV}$.  Similar coefficient values are obtained when trying to describe the $a_a(A)$-data in terms of the Thomas-Fermi (TF) theory\cite{Danielewicz:2003dd,Danielewicz:2004et}, represented in the figure.    The TF results suggest, nonetheless, that the effects of curvature of nuclear surface may, at some level, affect the symmetry energy even in the heaviest encountered nuclei.  It should be mentioned that, within the TF theory\cite{Danielewicz:2003dd,Danielewicz:2004et}, the local result~(\ref{eq:rhoa_loc}) for the isovector density is exact.

\begin{figure}[th]
\centerline{\psfig{file=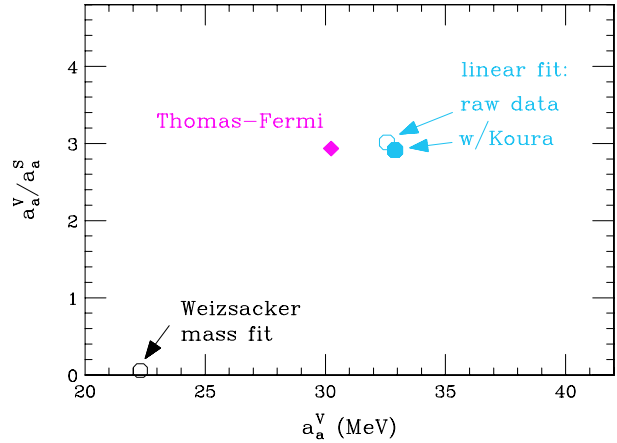,width=8cm,height=5.8cm} }
\vspace*{8pt}
\caption{Symmetry-energy parameter values in the plane of $a_a^V/a_a^S$ and $a_a^V$, from comparing different theoretical expectations to data.}
\label{fig:rabd}
\end{figure}

The parameters arrived at in different ways are further summarized in Fig.~\ref{fig:rabd}.  The favored values of symmetry energy at $\rho_0$ are $a_a^V = (30-33) \, {\rm MeV}$.  From Figs.~\ref{fig:rabd} and~\ref{fig:denscomp}, we read off the favored value of $L \sim 80 \, {\rm MeV}$.

Subsequent efforts in this investigation will be aimed at an extraction of the mass-dependent symmetry coefficients from spherical SHF calculations, at a model-independent understanding of the asymmetry skins, as well as at an understanding of the shell, Coulomb, deformation and curvature effects in the context of symmetry energy.

\section*{Acknowledgements}
This work was supported by the National Science Foundation under Grants PHY-0551164, PHY-0555893, PHY-0606007 and PHY-0800026.

\bibliography{coka08}

\bibliographystyle{myws}  

\end{document}